# High-Temperature Electromagnetic and Thermal Characteristics of Graphene Composites

Zahra Barani[1,2], Fariborz Kargar[1,2,*], Amirmahdi Mohammadzadeh[1,2], Sahar Naghibi[1], Carissa Lo[1], Brandon Rivera[2], and Alexander A. Balandin[1,2,*]

[1]Phonon Optimized Engineered Materials (POEM) Center, Materials Science and Engineering Program, University of California – Riverside, California 92521 USA

[2]Nano-Device Laboratory (NDL), Department of Electrical and Computer Engineering, Bourns College of Engineering, University of California, Riverside, California 92521 USA

[*] Corresponding authors: fkargar@engr.ucr.edu and balandin@ece.ucr.edu ; web-site: http://balandingroup.ucr.edu/

## Abstract

We describe a method for scalable synthesis of epoxy composites with graphene and few-layer graphene fillers, and report on the electromagnetic interference (EMI) shielding and thermal properties of such composites at elevated temperatures. The tested materials reveal excellent total EMI shielding of $SE_T \sim 65\,dB$ ($\sim 105\,dB$) at a thickness of $1\,mm$ ($\sim 2\,mm$) in the X-band frequency range of $f = 8.2\,GHz - 12.4\,GHz$. The room-temperature cross-plane thermal conductivity of the composite with $\sim 19.5\,vol\%$ of fillers was determined to be $K \approx 11.2 \pm 0.9\,Wm^{-1}K^{-1}$, which is a factor of $\times 41$ larger than that of the pristine epoxy. Interestingly, the EMI shielding efficiency improves further as the temperature increases to $520\,K$ while the thermal conductivity remains approximately constant. The excellent EMI shielding and heat conduction characteristics of such multifunctional graphene composites at elevated temperatures are promising for packaging applications of microwave components where EMI shielding and thermal management are important design considerations.

## I. Introduction

Rapid miniaturization and a consequent increase in the heat and electromagnetic wave (EM) emission in the densely-packed electronic systems make the simultaneous heat management and electromagnetic interference (EMI) shielding crucially important.[1–3] Any working electronic device is the source of EM radiation. For this reason, the electronic components have to be protected from the intra- and inter-system EM radiations in order to avoid fast degradation and failure.[4–14] With the universal deployment of wireless communications, portable devices, high-power transmission lines, the environmental EM pollution has become a major concern for human health as well. It has been suggested that prolonged exposure to even non-ionizing EM waves in the MHz and GHz frequency range may have detrimental effects on humans and other living beings.[15–19] The dense packing of the electronic components in the state-of-the-art 2D, 2.5D and 3D integrated systems and generation of high heat fluxes create an environment with elevated temperatures, which adversely affect the efficiency and stability of the EMI shielding.[20,21] The absorption of EM waves results in the temperature rise of the material making the situation even worse. The data on the efficiency of conventional and recent EM shield materials in most of cases are limited to room temperature (RT) operation.[20,21] The latter is despite the fact that many new non-metallic materials, introduced for EMI shielding, suffer from thermal instability, oxidation, or significant reduction in the shielding efficiency at high temperature. Evaluation of the performance of EMI shielding at high temperature is particularly important for certain high-power electronic systems operating in harsh environments.[20,21] These concerns require development of novel multifunctional materials, which can serve concurrently as an excellent EM shields with the high thermal stability at elevated temperatures. The ability of such materials to act as the thermal interface materials (TIM), which can dissipate heat efficiently becomes a necessity rather than an extra bonus feature.[2,3,20–22] There is a need for multifunctional materials which can perform both EMI shielding and TIM functions. The low-weight, mechanical stability, resistance to oxidation, flexibility, and ease of manufacturing are other important criteria that are considered to be the key parameters for multifunctional materials' applicability and cost-effective mass production. In this study, we synthesized and tested a multifunctional EMI shielding composite with graphene, which can efficiently function at elevated temperatures, above 500 K, without losing its EM shielding and heat conduction properties. In the context of present research, here and below, we use the term *graphene* for a mixture of graphene and few-layer graphene (FLG) flakes, which have nanometer-scale thickness and micrometer-scale lateral dimensions.

It is well known that the EMI shielding materials block the incident EM waves by reflection and absorption mechanisms. Both mechanisms require interaction of the EM waves with the charge carriers in the material. For this reason, EMI shielding materials should be either electrically conductive or contain electrically conductive fillers. Nonetheless, high electrical conductivity is not a requirement for EMI shielding and bulk resistivity on the order of 1 $\Omega$ cm is sufficient for the most industrial applications.[4,6,23] Recently, there was a growing demand for the EMI shielding materials capable of attenuating the microwaves at high temperatures, exceeding 500 K for applications in aerospace, defense, and high-power electronics. The conventional materials for EMI shielding are metals, which are utilized as coatings and enclosures.[20,23] Metals possess a high density of mobile charge carriers, which blocks the EM waves mostly by the reflection mechanism.[23] Metallic shields are heavy and prone to oxidation, resulting in corrosion. The electrical conductivity of metals and, correspondingly, shielding efficiency are degraded at elevated temperatures.[24,25] A new type of metallic fillers, MXenes, demonstrated high EMI shielding efficiency in the X-band frequency range.[4] However, these materials are prone to oxidation, particularly at increased temperatures, limiting their stability and EM shielding applications.[26,27]

An alternative approach to EMI shielding is the use of polymers laden with electrically conductive fillers.[23,28,29] The first generation of the EMI shielding polymers utilized large loading fractions of metallic particles. Although the polymer metal composites had reduced the problems associated with corrosion and oxidation, they suffered from heavy weight and degraded efficiency at elevated temperatures.[25,30] Ceramics,[31–38] carbon fibers,[39–46] carbon black,[47,48] carbon nanotubes,[49–55] graphite,[56–58] reduced graphene oxide,[5,9,59–69] graphene,[70–73] ferromagnetic materials,[74–77] and combinations of carbon allotropes with metallic and nonmetallic fillers[59,61,84,85,70,72,78–83] have been tested as potential fillers for EMI shielding. Ceramic fillers demonstrated a promise for high-temperature applications owing to their excellent thermal stability and relatively high thermal conductivity at high temperatures. However, ceramic materials are electrical insulators and do not absorb well the EM waves in the microwave spectral region.[20] Carbon allotropes such as graphene and carbon nanotubes demonstrated better performance as fillers in polymeric matrices. Other advantages of carbon fillers include the low weight, high thermal stability, and anti-corrosive properties.[20,55,86] From the fundamental science point of view, graphene is more compatible with the polymeric matrices than carbon nanotubes owing to its two-dimensional nature. Technologically, it graphene can be mixed and processed more conveniently without agglomeration and better coupling to the matrix material.

In EMI shielding applications at elevated temperatures, a high thermal conductivity of composites is not only an advantage but a requirement. The absorbed portion of EM wave energy by the EMI shield coating goes to heat. The heat has to be conducted away and dissipated. The reflection property of the EMI shielding material reduces at elevated temperatures because the electrical conductivity of metals decreases with increasing temperature. The high thermal conductivity of the composites is also beneficial for composites, which are used both as EMI shielding materials and TIMs. TIMs are materials which are applied between two solid surfaces in order to fill the microscopic voids at the interface and enhance the thermal transport from a heat source to a heat sink.[87–89] The base materials for TIMs are also amorphous polymers, which have low thermal conductivity, typically in the range from $0.2$ to $0.5 \, \mathrm{Wm^{-1}K^{-1}}$.[90] For this reason, the polymers used as base materials for TIM are filled with highly thermally conductive fillers to enhance its overall thermal transport. Among all fillers, mentioned previously, graphene and FLG have both high electrical conductivity and the highest thermal conductivity.[91–96] Composites filled with ceramics and other carbon allotropes exhibit lower thermal conductivity comparing to that of graphene. The in-plane thermal conductivity of MXene films is reported to be $\sim 3$ to $\sim 4 \, \mathrm{Wm^{-1}K^{-1}}$, which is an order of magnitude less than that of ceramics.[97]

The intrinsic thermal conductivity of graphene is in the range of $2000 - 5000 \, \mathrm{Wm^{-1}K^{-1}}$ depending on its lateral dimensions, crystal quality, and defect concentration.[94–96] For practical thermal applications, a mixture of graphene and FLG is the most beneficial because FLG retains the excellent heat conduction properties while offering higher cross-section area for the heat flux. Owing to their mechanical flexibility, graphene fillers facilitate the formation of the interconnected network of FLG fillers. Phonons are the main heat carriers in graphene and FLG. For this reason, large lateral dimensions of the fillers are required in order to preserve the heat conduction properties.[98] The phonon mean free path (MFP) in graphene is $\sim 750 - 800$ nm, suggesting that the fillers used in composites should have lateral dimensions exceeding the phonon MFP.[99] The thermal transport properties of FLG are expected to be less affected by the exposure to the matrix material. In our recent study, we reported an enhancement of the thermal conductivity by a factor of $\times 51$ at $\phi = 45$ vol% loading fraction of a mixture of graphene and FLG.[88] The thermal conductivity of $\sim 15$ $\mathrm{Wm^{-1}K^{-1}}$ can now be routinely reached by an addition of graphene fillers to polymer matrices.[100–102] The thermal conductivity enhancement factor depended on the quality of the filler, its average lateral dimension, and the composite synthesis procedure.

Despite the importance of knowing EMI shielding efficiency at elevated temperatures and thermal properties of EMI composites there have been few studies that provided such data. Here, we report on the EMI shielding properties and thermal characteristics of the multifunctional epoxy – based composites with FLG fillers. The composites exhibit room temperature cross-plane thermal conductivity of $11.2 \pm 0.9$ $\mathrm{Wm^{-1}K^{-1}}$ and the total shielding efficiency of 65 dB at a thickness of $t = 1$ mm at the filler loading of $\phi = 19.5$ vol% in the X-band frequency range. The EMI shielding and thermal properties of the samples were examined at the temperatures beyond 500 K. The results show that the graphene – epoxy samples preserve their excellent heat conduction properties at high temperatures while the EMI shielding efficiency even improves. The excellent EMI shielding and thermal properties of the multi-functional graphene composites at elevated temperatures along with their thermal stability and light-weight are important for electronic packaging and airborne systems where efficient EMI shielding, low cost, and low weight are required at high temperatures and high power densities.

## II. Material Synthesis and Characterization

The materials synthesis processes and quality control steps for epoxy-based composites with graphene and FLG fillers are illustrated in Figure 1 (a-e). The samples were synthesized with the commercially available graphene powder (xGnP, grade H, XG-Sciences, USA) and epoxy (Allied High-Tech Products, Inc.). It has been shown that graphene fillers with a larger aspect ratio provide the highest EMI shielding efficiency and thermal conductivity.[2,88,103–107] This explains the selection of the graphene powder with the largest average lateral dimensions of the flakes. A representative scanning electron microscopy (SEM) image of FLG utilized in the composite preparation confirming their large lateral dimensions in the range of $15 - 25$ µm is shown in Supporting Information Figure S1 (a-b). In order to prepare composites with different filler concentrations, FLG powder was added to epoxy resin in certain amounts, followed by the addition of the hardener. The compound was mixed using a high shear speed mixer (Flacktek Inc) in several steps to obtain a uniform dispersion. The mixture was vacuumed several times at least for 15 minutes at each step in order to remove any possible voids or air bubbles. The final mixture was poured into round molds, gently pressed, and left to solidify at RT for approximately 8 hours. After curing, the samples were placed to a furnace for ~1 hour at 120 ℃ (Figure 1 a-b). It should be noted that in the EMI shielding experiments, the thickness of the sample affects its absorption and thus, its total shielding efficiency. Therefore, for detailed comparison, all samples were polished down to the same thickness of ~1.00 mm. An optical image of the pristine epoxy and the composite with 11.4 vol% and 15.3 vol% of graphene fillers is shown in Figure 1 (c).

[Figure 1]

The quality of the samples was verified by the SEM inspection and Raman spectroscopy. Raman spectra of representative samples with different graphene loadings are shown in Figure 1 (d). The intensity of the characteristic Raman peaks of graphene, namely the G peak at ~$1570$ cm$^{-1}$ and 2D band at $2700$ cm$^{-1}$, increases as more graphene is loaded to the epoxy. These two peaks are not distinguishable in the samples with graphene loading less than $\phi \leq 1.5$ vol%. The variations in the graphene filler concentration and their overlap are clearly seen in the SEM image of the composite with 15.3 vol% of the fillers (see Figure 1 (e)). Additional data of SEM inspection of the samples are provided in the Supporting Information (Figure S1 (c-f)). The data confirm that at high loadings, graphene fillers overlap forming the electrically percolated networks. The filler attachments and percolation strongly affects the electrical and thermal characteristics of the composites, and finally their EMI shielding performance at RT and above. The process of EM wave interaction with the graphene composite is illustrated in Figure 1 (f). Large hexagons indicate FLG fillers. A fraction of the incident EM wave is reflected at the air-composite interface. A part of the EM wave, which passes through the composite's interface, is absorbed directly or after multiple internal reflections inside the composite. Only a small fraction of the incident EM wave is transmitted through the samples. The absorption of EM waves results in the increase of the composite's temperature, which explains the need for high thermal conductivity of EMI shielding material.  In the schematic, the red arrows indicate the highly conductive paths for heat dissipation through the fillers inside the epoxy polymer base. The EMI shielding efficiency and the cross-plane thermal conductivity of the samples were measured using a programmable network analyzer (see Figure S2) and a "laser flash" instrument (see Figure S3). The schematic and operational principles of both systems are shown in Figures 1 (g-h). The details of the experimental setups and measurements are provided in the Methods and Supporting Information.

## III. Results and Discussion

**EMI Shielding:** To determine EMI characteristics, we measured the scattering parameters, $S_{ij}$, using the two-port PNA system at RT and higher temperatures. The indices $i$ and $j$ represents the ports, which are receiving and emitting the EM waves. Each port can simultaneously emit and detect the EM waves and thus, the results of the measurements include four parameters of $S_{11}$, $S_{12}$, $S_{21}$, and $S_{22}$. Given that the samples are symmetric, one can expect that $|S_{11}| = |S_{22}|$ and $|S_{12}| = |S_{21}|$. The scattering parameters are related to the coefficients of reflection, $R = |S_{11}|^2$, and transmission, $T = |S_{21}|^2$.[108] Knowing these two coefficients, one can obtain the coefficient of absorption, $A$, according to $A = 1 - R - T$. This is because for any EM wave incident on the sample, a fraction is reflected at the interface. A part of EM wave, which enters the composite, is absorbed and the rest is transmitted. The coefficient of absorption, which is defined as the power percentile of the absorbed EM way in the medium to the total power of incident wave, is not truly indicative of material's ability in absorbing the EM waves. This is because some part of the wave is reflected at the interface prior to entering the medium. For this reason, the effective absorption coefficient, $A_{eff}$, is defined as $A_{eff} = 1 - R - T/1 - R$. After determining $R$, $T$, and $A_{eff}$, the shielding efficiency of the material can be calculated.

Figure 2 (a) shows the coefficients of $R$, $A_{eff}$, and $T$ of the epoxy composites as a function of graphene loading fraction at a representative frequency of $f = 9$ GHz. The main observation is that addition of even small loading of graphene results in rapid increase in the reflection and affective absorption, and a corresponding decrease in transmission. For the filler concentrations greater than $\phi > 2.5$ vol%, the rate of change for all three coefficients becomes small. The reflection increases from ~10% for pristine epoxy to ~85% for the samples with $\phi \geq 4.9\%$ and remains constant with negligible variations at higher loading fractions. The effective absorption exhibits a similar trend, showing an abrupt increase from 42% at $\phi = 2.5$ vol% to 80% at $\phi = 4.9\%$, and an asymptotical approach to 100% at the higher loadings. At a high loading of $\phi = 19.5$ vol%, the blocking of EM waves by reflection and absorption becomes so strong that only 0.00004% of the incident wave is transmitted through the medium (Figure 2 (a), black triangles). The calculated $R, T, A$ and $A_{eff}$ as a function of the frequency for the samples with different FLG concentrations are presented in Supplementary Information Figure S4.

The attenuation of EM waves in a conductive medium is related to the skin depth, $\delta$.[108] In good conductors, $\delta = (\pi f \mu \sigma)^{1/2}$ where $f, \mu$, and $\sigma$ are the EM wave frequency, medium's permeability and electrical conductivity, respectively. Graphene and FLG are good conductors of electricity. The sheet resistance of SLG and FLG vary from ~100 $\Omega$ up to 30 k$\Omega$ depending on the number of layers and quality.[91–93] Pristine epoxy, on the other hand, is an electrical insulator with the conductivity in the order of ~$10^{-16}$ Sm$^{-1}$.[109] Addition of FLG to the epoxy, increases the composite's electrical conductivity after the establishment of the percolation network of the fillers. The results of the standard four-probe in-plane electrical conductivity measurements of the epoxy with FLG as a function of the filler loading are presented in Figure 2 (b). The details of the measurements are explained in the Methods and Supporting Information. The abrupt change in conductivity observed at $\phi = 2.5$ vol% indicates the onset of the electrical percolation where the large FLG fillers inside the insulating polymer matrix create an electrically conductive network. Electrical percolation is theoretically described by power scaling law, $\sigma \sim (\phi - \phi_{th})^t$, where $\phi_{th}$ is the filler loading fraction at the electrical percolation threshold, and $t$ is the "universal" critical exponent.[103,110] The coincidence of the electrcial percolation threshold shown in Figure 2 (b) and the filler loading at which the abrupt change in the effective absorption and transmission is observed demonstrate the correlation among these parameters.

[Figure 2]

The total shielding efficiency, $SE_{tot}$, describes the total attenuation of the incident EM wave as it hits and passes through the composite. This parameter determines the material's ability to block the EM waves and consists of two terms of reflection loss, or reflection shielding efficiency, $SE_R$, and absorption loss, or absorption shielding efficiency, $SE_A$. These parameters can be calculated in terms of $R$, $T$, and $A_{eff}$ as follows[2,108]

$$SE_R = 10 \log[(P_i/(P_i - P_r)] = -10\log(1 - R), \qquad (1)$$

$$SE_A = 10 \log[(P_i - P_r)/(P_i - P_r - P_a)] = 10 \log(1 - A_{eff}), \qquad (2)$$

$$SE_T = 10 \log(P_i/P_t) = SE_R + SE_A, \qquad (3)$$

where $P_i$, $P_r$, $P_a$, and $P_t$ are the power of incident, reflected, absorbed, and transmitted EM wave, respectively.

Figure 3 (a-c) shows $SE_R$, $SE_A$, and $SE_T$ of the samples as the functions of the graphene filler loading. As more fillers are loaded to the pristine epoxy polymer, $SE_R$ rapidly increases (Figure 3(a)). At $\phi \geq 2.5$ vol%, the enhancement in reflection becomes slower, and the $SE_R$ saturates, *i.e.* it does not increase significantly with more graphene loading. The reflection efficiency is directly related to the surface electrical conductivity or, in other words, the concentration of free charge carriers at or near the interface. The saturation behavior of $SE_R$ indicates that the large graphene fillers are overlapping at the surface of the sample, increasing the overall electrical conductivity of the sample by orders of magnitude as compared to that of pristine epoxy or composites with low loading. Note that at $\phi = 2.5$ vol%, $R\sim73\%$ indicating that most part of the incident EM wave is already reflected back at the interface. In contrast, $SE_A$ increases with adding more fillers (Figure 3 (b)). In samples with $\phi \geq 2.5$ vol%, the absorption loss becomes the dominant shielding mechanism. This can be seen clearly at higher filler loadings where $SE_A$ approaches 52 dB at $\phi = 19.5$ vol% meaning that almost all the remaining fraction of the EM wave, which passes through the medium after reflection, is being absorbed by the composites. Figure 3 (c) presents the total EMI shielding efficiency of the samples as a function of the graphene loading. The average reflection, absorption, and total shielding efficiency of composites over the complete X-band frequency range as a function of graphene loading are shown in Figure 3 (d). As it is seen and described previously, the reflection loss (red circles) initially increases but then saturates and becomes almost constant at $\phi \geq 2.5$ vol% whereas the absorption and thus, total shielding efficiency continue to increase with addition of more fillers. One can summaries the difference in behavior by noting that reflection is mostly a surface phenomenon while absorption happens ij the volume of the composite. The electrically conducting percolation network forms at the surface at lower graphene loading than in the volume of the material.

[Figure 3]

In Figure 4 (a-b) we present the EMI shielding efficiency as a function of the thickness for the epoxy composites with 17.1 vol% and 18.8 vol% graphene loading, respectively. The absorption shielding efficiency is directly related to the thickness of the sample. The classical Simon's equation relates the $SE_A$ to the thickness according to $SE_A = (1.7t/\rho)f^{0.5}$ where $t$ [cm] and $\rho$ [$\Omega$cm] are the sample's thickness and bulk DC resistivity and $f$ [MHz] is the frequency. With the increasing thickness, the $SE_A$ increases, which results in the enhancement of the total EMI shielding efficiency (see Eq. (3)). The dependence of $SE_T$ as a function of the sample's thickness is shown in Figures 4 (c-d) for the composites with 17.1 vol% and 18.8 vol% at the fixed frequencies of 8.5 GHz, 9.5 GHz, and 11.5 GHz. The dashed lines are the fitted linear regression over the experimental data. A clear linear trend is observable in both panels. It should be noted that for the epoxy composite with graphene loading of 18.8 vol% and thickness of ~2 mm, the total shielding efficiency at different frequencies is between 95 dB to 105 dB. The values achieved in our samples exceed the EMI shielding requirements for most industrial applications, which are at the level of the total 30 dB shielding efficiency.[2,23] Another advantage of the graphene-enhanced epoxy composites is that they provide higher EM shielding at much lower weights and with stable mechanical structure as compared to conventional metallic enclosures, which are widely used in electronic packaging.

[Figure 4]

Densely packed electronics, in addition to emitting undesirable EM waves to the environment, dissipate high amounts of heat, which leads to the increase in temperature of all components in the electronic package. A part of the EM wave, which turns into heat during the absorption process, can be described by the following equation:[20]

$$P = \frac{1}{2}\sigma|E|^2 + \pi f \varepsilon_0 \varepsilon_r''|E|^2 + \pi f \mu_0 \mu_r''|H|^2. \tag{4}$$

Here, $P$ is the power, $\varepsilon_0$ and $\varepsilon_r''$ are the vacuum and imaginary part of medium's relative permittivity, and $\mu_0$ and $\mu_r''$ are the vacuum and imaginary part of medium's relative permeability. The first, second, and the third terms in the right hand side correspond to the Joule heat loss, dielectric loss, and magnetic loss, where the latter is negligible for the epoxy composites with graphene. The temperature-rise in most cases adversely affect the ability of materials in blocking the EM waves. In this regard, the EM shield material must possess a high thermal stability, *i.e.* preserving its EM shielding characteristics even at elevated temperatures at least up to 500 K, and relatively high thermal conductivity in order to dissipate heat to the environment.

Figure 5 (a-c) shows the EM shielding efficiency of three composites with 11.4 vol%, 13.4 vol%, and 17.1 vol% graphene loading at four selected temperatures, in the temperature range from 298 K to 528 K. The total shielding efficiency of the samples enhances by $\sim 8$ dB for all filler loadings as the temperature increases from RT to $\sim 520$ K. In Figure 5 (d-f), we present color maps of $SE_T$ as the function of temperature and frequency for composites with the low (3.5 vol%), medium (9.3 vol%), and high (17.1 vol%) graphene loadings. In contrast to some prior studies, we observed a continuous increase in the $SE_T$ for all filler loadings.[30,111] The increase in the $SE_T$ is originated mainly from the enhancement in absorption loss rather than increase in the reflection attenuation (see Figure S5). The epoxy composites reach the electrical percolation threshold at a rather low FLG concentration of $\phi \sim 2.5$ vol%. In these samples, the absorption is defined by the electrical conductivity. In a disordered system of randomly dispersed FLG fillers in a polymer matrix, the conductivity by migrating electrons ($\sigma_m$) in the plane of FLG fillers and conductivity by hopping electrons ($\sigma_h$) among the different FLG fillers affect the bulk average electrical conductivity of the composite. The hopping conductance mechanism is generally described as $\sigma_h \sim e^{-(T_0/T)^p}$ where $T_0$ depends on the material's property and $p$ is a parameter defined by the type of hopping. As the temperature increases, the hopping conductivity strongly increases. One should note that it is known that the electrical conductivity in graphite and graphene also increases with rise.[111,112] For these reasons, both mechanisms of electrical conduction in epoxy-graphene composites enhance with increasing temperature. The latter explains the increase in the absorption and the EMI shielding efficiency at elevated temperatures.

The high thermal stability of epoxy composites laden with FLG fillers provides several advantages over other types of EM shielding materials and fillers, *e.g.* ceramics, ferromagnets, or MXenes. Ceramic-based composites usually use SiC, which, although thermally stable, has rather poor absorption properties. The ferromagnetic fillers lose their magnetic properties as the temperature passes the Curie temperature, causing the drop in their magnetic shielding characteristics at elevated temperatures. The MXene fillers and films exhibit high shielding efficiencies but prone to formation of oxide layers and reveal poor thermal conductivity making them unsuitable for high-temperature EMI shielding applications.[26,27,97,113]

[Figure5]

**Thermal Characteristics:** The cross-plane thermal diffusivity and conductivity of the composites at RT and at elevated temperatures were measured using the "laser flash" analysis (LFA, 467 HyperFlash, NETZSCH, Germany). The schematic and operational principal of the measurements are shown in Figures 1 (i). The details of the experimental setup and measurements are provided in the Methods. The thermal diffusivity, $\alpha$, was measured directly while the thermal conductivity, $K$, was calculated as $K = \rho \alpha c_p$, where $\rho$ and $c_p$ are the mass density and the specific heat. Figure 6 (a-b) shows the cross-plane thermal diffusivity and thermal conductivity of the composites as the functions of the graphene loading. Both thermal characteristics, $\alpha$ and $K$, grow fast with increasing graphene loading. The super-linear, nearly of higher than quadratic, dependence of the thermal characteristics of the composites as the function of the filler loading is attributed to creation of the highly thermally conductive paths of the percolated FLG fillers. The mechanism of the thermal percolation in epoxy composites with graphene has been described by us previously.[88,100,114] One should note that epoxy composites with only 19.5 vol% FLG loading provide cross-plane thermal conductivity of $11.2 \pm 0.9 \, \mathrm{Wm^{-1}K^{-1}}$, which is comparable to that of the ceramic materials, widely used in high–temperature EMI shielding, specifically for their rather high thermal conductivity and thermal stability. The advantage of graphene composites over conventional ceramics is that their EMI shielding efficiency is much higher and their weigh is much lower than that of conventional ceramics.

In order to investigate the thermal stability of the composites, we carried out temperature dependent thermal diffusivity and conductivity experiments in the temperature range between RT to 460 K. the results of the measurements are presented in Figures 6 (c-d). The thermal diffusivity of the samples decreases with increasing the temperature. The rate of change is more significant for the composites with higher graphene loadings. However, the thermal conductivity is almost constant owing to the increase in the specific heat as the temperature increases. One should note that the reported data is for the thermal transport properties in the cross-plane direction. The in-plane thermal conductivity of the samples is expected to be much higher because graphene flakes tend to align in the in-plane direction during the fabrication process.[102] The nearly constant high thermal conductivity of the composites over a wide temperature range demonstrates that epoxy with graphene fillers can serve as a multifunctional composite for applications in electronic systems, providing excellent EMI shielding, heat dissipation, mechanical support and adhesion even at elevated temperatures.

[Figure 6]

## IV. Conclusions

We examined the EMI shielding efficiency and thermal conductivity of multi-functional epoxy composites with graphene fillers at elevated temperatures. It was found that graphene composites with the filler loading of $\phi = 19.5$ vol% exhibit a high total shielding efficiency of 65 dB (105 dB) at a thickness of 1 mm (2 mm) in the X-band frequency range. The same composite reveals high thermal conductivity of $11.2 \pm 0.9 \, \mathrm{Wm^{-1}K^{-1}}$ at RT. The composites preserve their high thermal conductivity at elevated temperatures while their EMI shielding efficiency increases for all examined filler loadings. The excellent EMI shielding and heat conduction properties of graphene composites are important for the thermal management and electromagnetic compatibility of the next generation of high-power, high-temperature electronic devices where the weight, cost, and functionality at harsh environments are of utmost importance.

## METHODS

**Sample Preparation:** The samples were prepared by mixing the commercially available FLG flakes (xGnP graphene nanoplatelets, grade H, XG-Sciences) with epoxy (Allied High Tech Products, Inc.). For samples with less than 15 vol% loading fraction of FLG, the epoxy resin and the filler were mixed using a high-shear speed mixer (Flacktek Inc.) at 800 rpm and 2000 rpm each for 5 minutes. The mixture was vacuumed for 10 minutes at least three times. At the end, the curing agent (Allied High Tech Products, Inc.) was added in the mass ratio of 12:100 with respect to the epoxy resin. The mixture was mixed and vacuumed one more time and left in the room temperature for 8 hours to be cured. The prepared composite samples are put in the furnace for ~2 hours at 120º C in order solidify better. For samples with more than 15 vol% loading fraction of FLG, there were two additional steps. First, the solution was mixed at high rotation speeds of 3000 for 10 to15 seconds while adding graphene to the resin at the beginning of the procedure. At the end, the homogenous mixture was gently pressed before being left to cure. The details of the sample preparation are provided in the Supplementary Materials.

**Electrical Conductivity Measurements:** The in-plane DC resistance and conductivity of the graphene composites with the FLG fillers using standard 4-probe configuration. For this purpose, four needles contacted the surface of the sample using the probe station. The schematic and experimental setup of the measurements are shown in Supporting Information Figure S7. The distance between the contacts was $\sim$5 mm. The current was applied to the outer contacts and the voltage difference was measured between the two inner contacts. The resistivity was calculated, according to $\rho = GV/I$, where $G$ is the geometrical factor and $V$ and $I$ are the measured voltage and applied current. The geometrical factor can be calculated according to $G = \frac{\pi}{Ln2} t T_2 \frac{t}{s}$ where $t$ is the thickness of the samples and $s$ is the distance between two adjacent contacts. The coefficient $T_2$ is the correction factor extracted from the standard plots. If $\frac{t}{s} \ll 1$, one can assume that $T_2 t/s \rightarrow 1$ and thus, $G = 4.5324t$.[115] The conductivity is calculated as $\sigma = 1/\rho$.

**Electromagnetic Interference Shielding Measurements:** RF measurements were performed in the X-Band frequency range (8.2 – 12.4 GHz) with frequency resolution of 3 MHz at RT and elevated temperatures. A Programmable Network Analyzer (PNA) Keysight N5221A was used. The PNA was calibrated for 2-port measurements of scattering parameters $S_{ij}$ at input power $P_{in} = 3$ dBm. A WR-90 commercial grade straight waveguide with two adapters at both ends with SMA coaxial ports was used as a sample holder. In order to carry out high-temperature RF measurements, the waveguide with the sample were heated using a hot-plate heater. The temperature of the waveguide and the samples were measured using 6 thermocouples attached at different points of the waveguide (see Figure 1 (g)). Prior to collecting the temperature dependent RF measurements, the waveguide without any samples was tested in the temperature range between RT and 520 K. The performance of the waveguide did not change with temperature rise. Figure S6 shows the $SE_T$ of the bare waveguide without any samples as a function of temperature. Special cables were used for high temperature RF measurements. The samples with diameter $d \geq 25$ mm were a bit larger than the rectangular cross section ($22.8 \times 10.1$ mm$^2$) of the central hollow part of the waveguide in order to prevent the leakage of EM waves from the sender to receiver antenna. The scattering parameters, $S_{ij}$, were directly measured and used to extract the reflection and absorption shielding efficiency of the composites.

**Mass Density Measurements:** According to Archimedes principle and by using an electronic scale (Mettler Toledo), the mass density of the samples was measured. In this method the density is defined using $\rho_c = (w_a/(w_a - w_w)) \times (\rho_w - \rho_a) + \rho_a$ formula where $w_a, w_w$ are the sample's weight in air and in water and $\rho_w$ and $\rho_a$ are the density of the ionized water and air ($0.0012$ gcm$^{-3}$) at room temperature.

**Thermal Diffusivity, Specific Heat Capacity, and Thermal Conductivity Measurements:** Thermal diffusivity of the samples were measured using the transient "laser flash" technique (LFA 467 HyperFlash, NETZSCH) compliant with the international standards of ASTM E-1461, DIM

EN 821 and DIN 30905. The heat capacity is calculated based on comparison of the temperature rise of the sample with that of the known reference sample. Later on the values off thermal diffusivity obtained directly from LFA, were used to determine the thermal conductivity value of the samples. Knowing the density, thermal diffusivity and the specific heat of the sample, thermal conductivity was calculated according to the equation $K = \rho \alpha c_p$ where $K$, $\rho$, and $c_p$ are the thermal conductivity, density, and specific heat, respectively. In LFA technique, the front surface of a plane-parallel sample is heated by a short energy pulse. From the resulting temperature excursion of the rear face measured with an infrared (IR) detector, the thermal diffusivity is obtained. More details on the LFA thermal conductivity method can be found in Supplementary Materials.

### Acknowledgements

The work at UCR was supported, in part, by the Office of Technology Partnerships (OTP), University of California via the Proof of Concept (POC) project "Graphene Thermal Interface Materials." A.A.B. also acknowledges the UC - National Laboratory Collaborative Research and Training Program - University of California Research Initiatives LFR-17-477237.

### Contributions

A.A.B., F.K. and Z.B. conceived the idea of the study. A.A.B. and F.K. coordinated the project and contributed to the experimental and theoretical data analysis. Z.B. prepared the composites, performed thermal and EMI shielding measurements and conducted data analysis. A.M. contributed to the sample characterization; S.N., C.L., and B.R. assisted with sample preparation; Z.B. and F.K led the manuscript preparation. All authors contributed to writing and editing of the manuscript. The authors acknowledge useful discussions on EM measurements with Dr. Alexander Khitun (UC Riverside) and thank Alec Balandin (Riverside STEM Academy) for assistance with the composite preparations and Raman measurements.

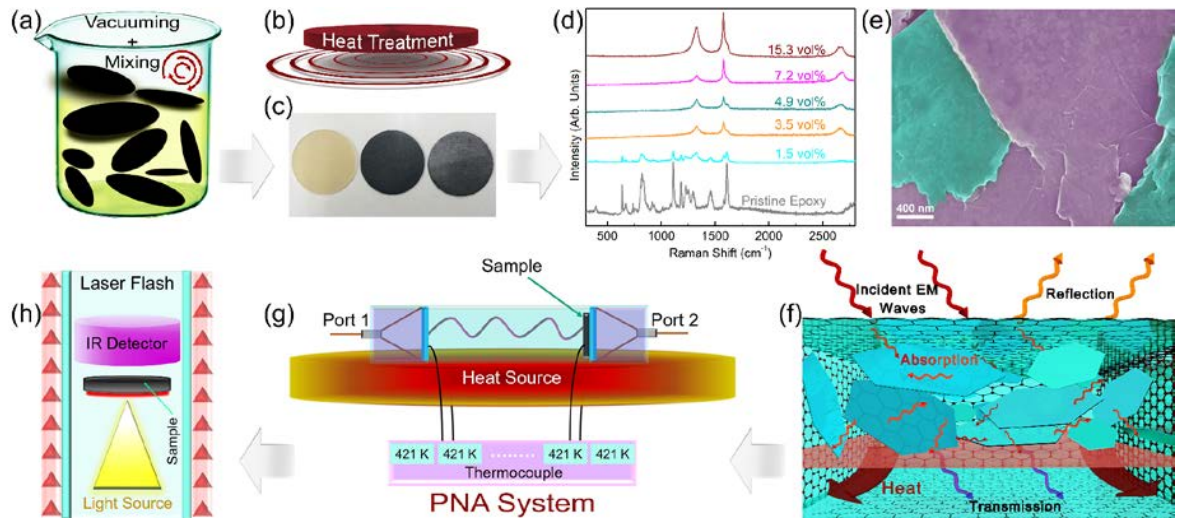

**Figure 1:** Process flow for the sample preparation and characterization. (a) Mixing and vacuuming FLG with the epoxy resin and hardener. The mixture is left to solidify for ~8 hours at RT. (b) Heat treatment of the cured samples at 120 ℃ for ~1 hour. (c - left to right) Optical image of a pristine epoxy sample, and an epoxy sample with 10.5 vol% and 17.6 vol% loading of FLG. (d) Raman spectra of epoxy with various FLG filler loadings. (e) SEM image of a composite with 17.6 vol% of FLG fillers, revealing filler overlapping inside the matrix. (f) Schematic showing the interaction of the EM waves with the composite. (g) Schematic of the experimental setups for the temperature dependent EMI shielding measurements, and (h) thermal conductivity measurements.

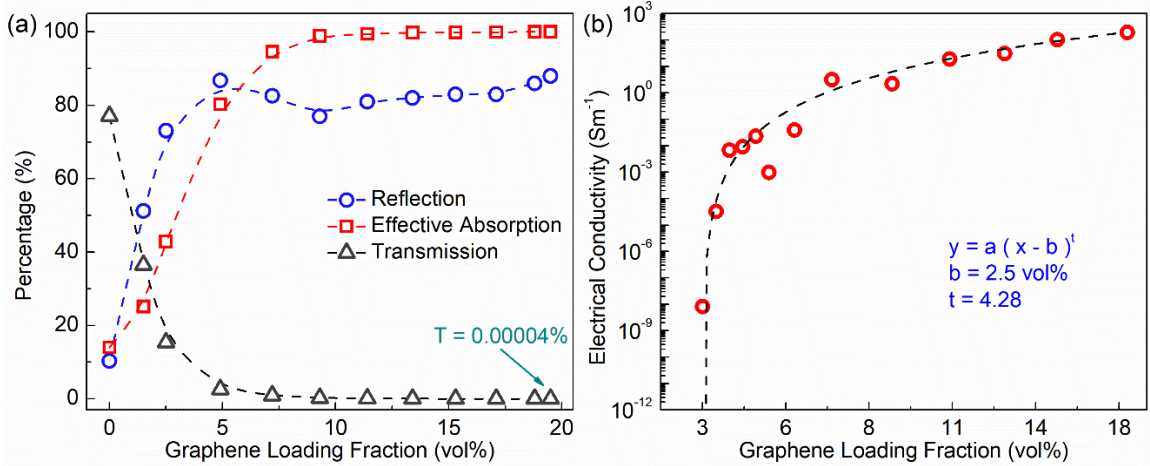

**Figure 2:** (a) Coefficients of reflection ($R$), effective absorption ($A_{eff}$), and transmission ($T$) as a function of filler loading at $f = 9.0$ GHz. The rate of variation of $R$, $A_{eff}$, and $T$ becomes smaller for $\phi > 2.5$ vol%. (b) Bulk in-plane electrical conductivity versus filler loading. The electrical percolation threshold is extracted at $\phi_{th} = 2.5$ vol% *via* fitting the experimental data using $\sigma \sim (\phi - \phi_{th})^t$.

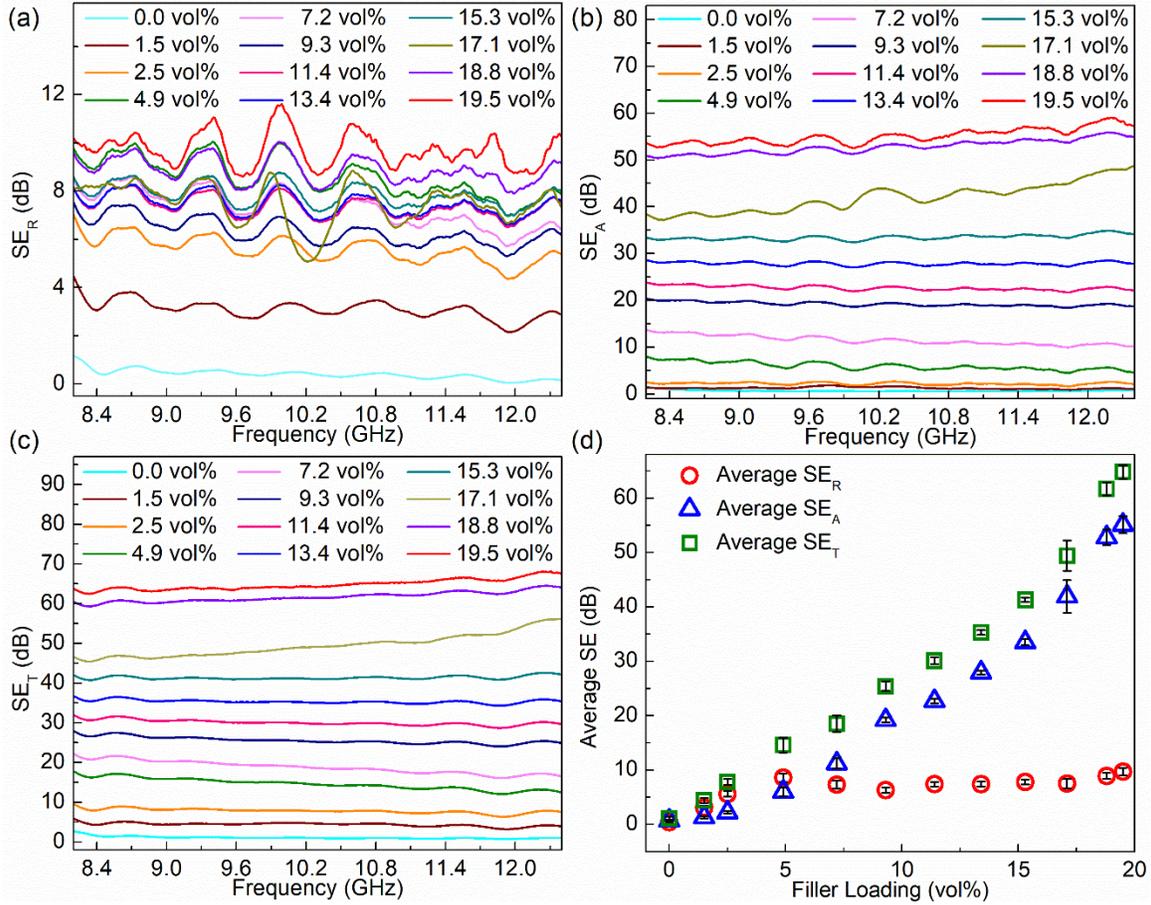

**Figure 3:** (a) Reflection ($SE_R$), (b) absorption ($SE_A$), and (c) total EMI shielding efficiency ($SE_T$) of the composite of 1 mm thickness in the frequency range from 8.0 GHz to 12.4 GHz at different graphene loading fractions. With the increasing graphene loading, the EMI shielding efficiency of the composites improves significantly. The total EMI shielding of the composites reaches to ~65 dB at 19.5 vol% of the graphene filler concentration.

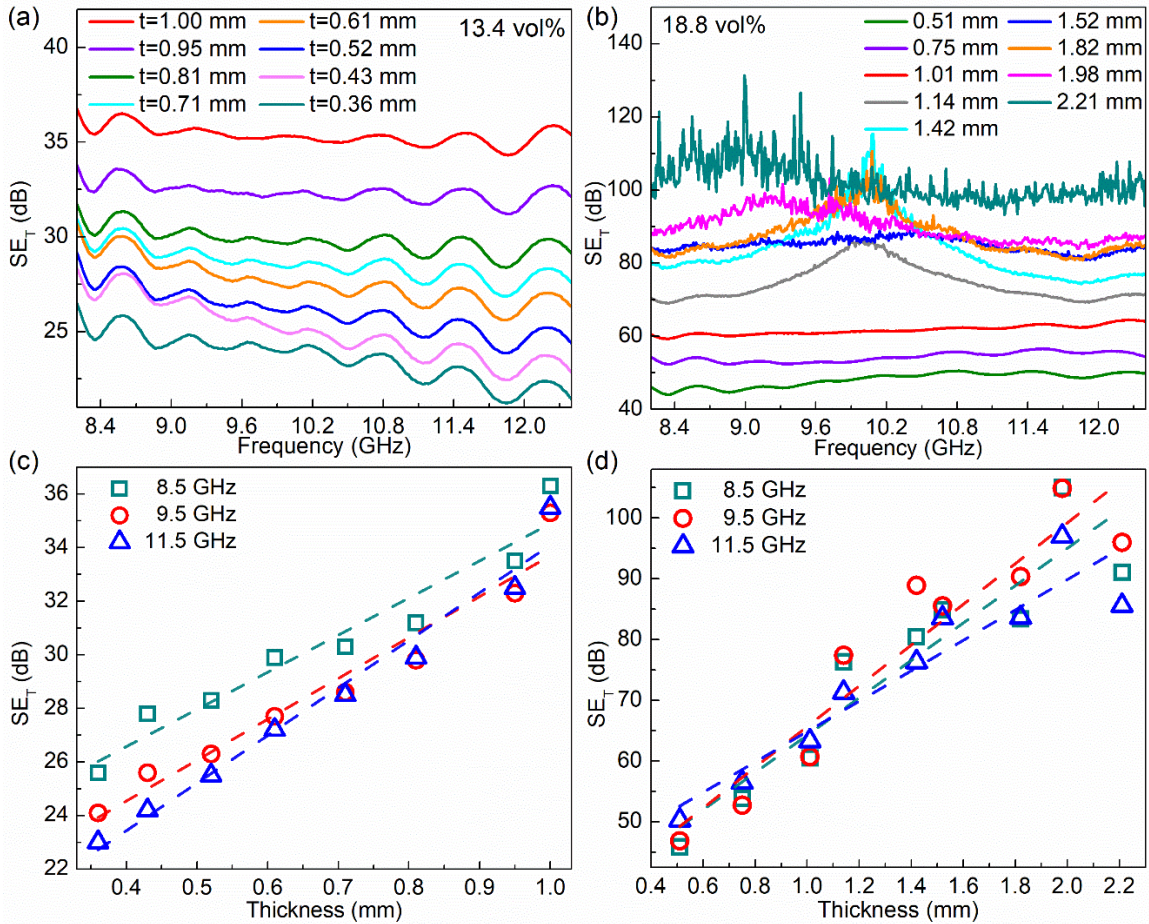

**Figure 4:** Total EMI shielding of epoxy composites with (a) 13.4 vol% and (b) 18.8 vol% graphene filler loading as a function of frequency at different thicknesses. Total EMI shielding of the epoxy composites with (c) 13.4 vol% and (d) 18.8 vol% loading of the filler versus thje thickness of the composites at frequencies of 8.5, 9.5, and 11.5 GHz. Note that the EM shielding increases linearly with increasing the thickness of the composite owing to the increase in shielding by absorption.

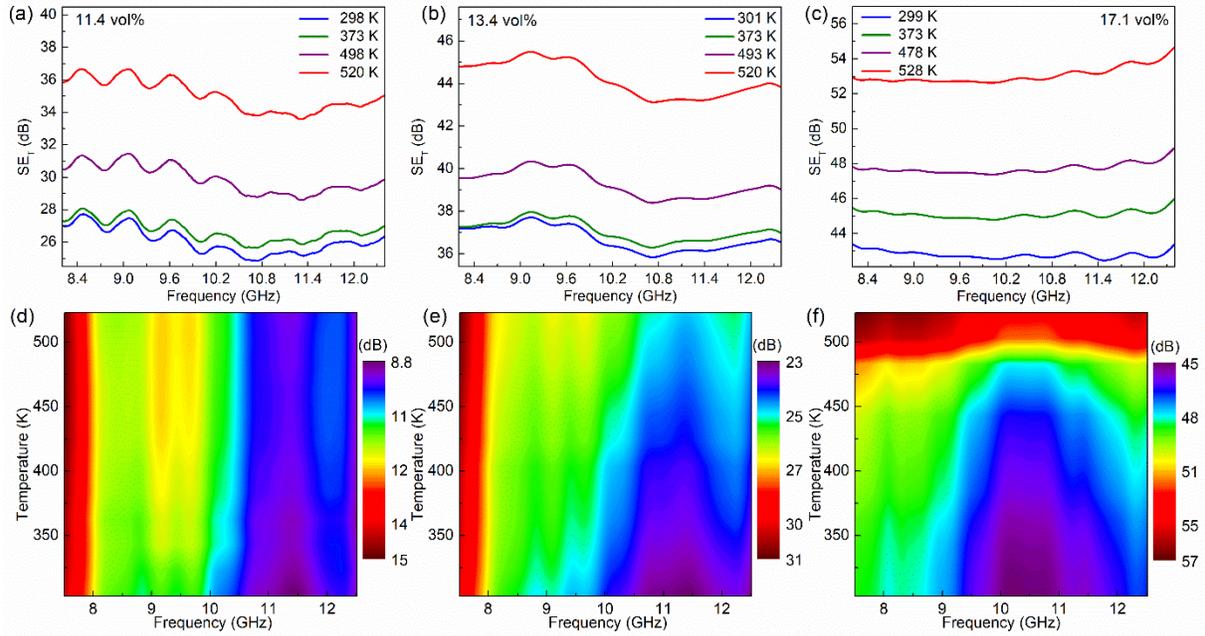

**Figure 5:** Total EMI shielding efficiency of the epoxy with (a) 11.4 vol%, (b) 13.4 vol% and, (c) 17.1 vol% of graphene fillers as a function of temperature in the X-band frequency range. The EMI efficiency increases with temperature in all samples. The color map of the total EMI shielding efficiency of the composites with (a) 3.5 vol% (b) 9.3 vol% and (c) 17.1 vol% of graphene filler loading as a function of frequency in the temperature range from 303 K to 523 K. The EMI attenuation for all samples increase as the temperature rises. The enhancement is more pronounced at higher frequencies.

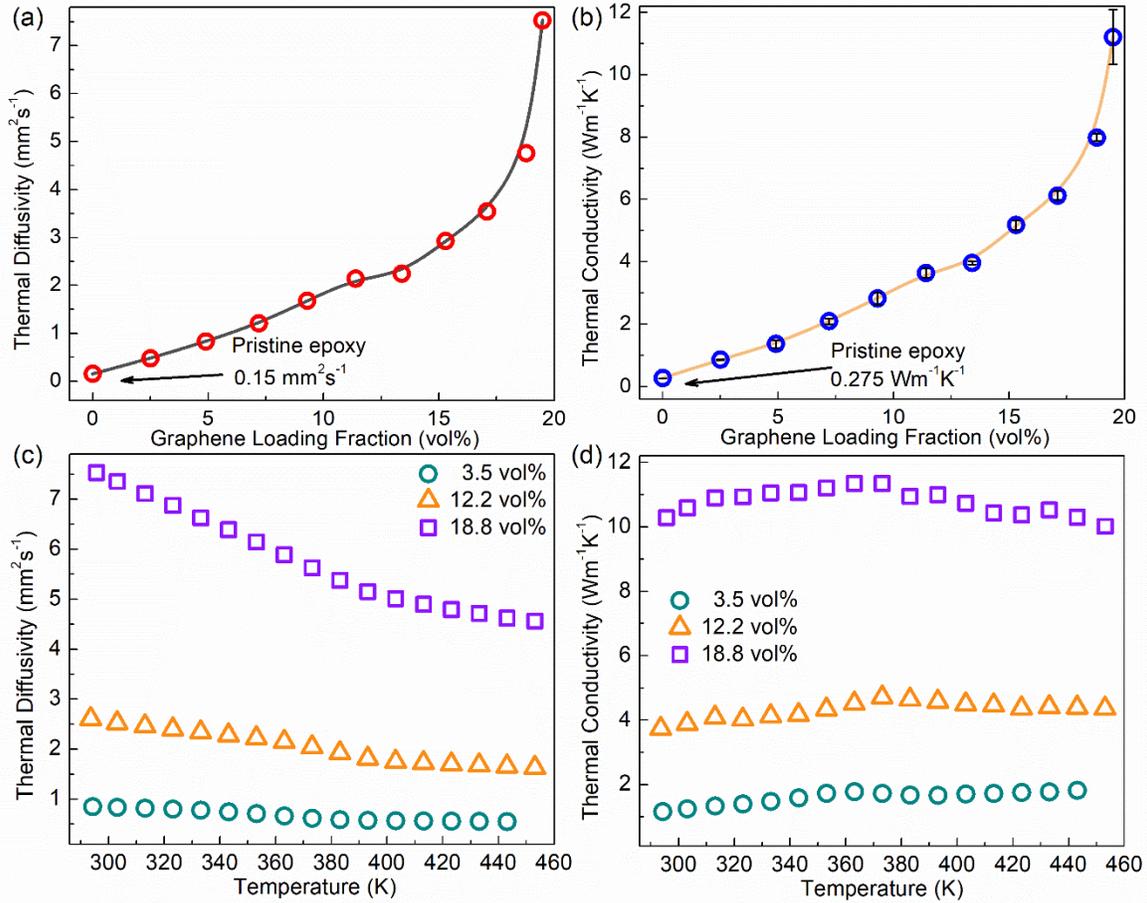

**Figure 6:** Thermal properties of the graphene composites. (a) Thermal diffusivity and (b) thermal conductivity of the composites as a function of graphene loading fraction at room temperature. The thermal conductivity exhibits a quadratic dependence on the filler loading except near the loading $\phi \sim 11.4$ vol% where there is plateau-like dependence attributed to the formation of graphene clusters. (c) Thermal diffusivity and (d) thermal conductivity of three different composites with the low, medium, and high loading of graphene fillers as a function of temperature. The thermal diffusivity decreases with temperature. The thermal conductivity remains nearly constant because the increase in the specific heat compensates for the decrease in thermal diffusivity as the temperature rises.